\documentclass[reprint,amsmath,amssymb,aps,prab,floatfix]{revtex4-2}
\usepackage[english] {babel}
\usepackage{amsfonts,graphicx,soul,natbib,mathrsfs}
\usepackage[euler]{textgreek}
\usepackage{epstopdf}
\usepackage{xcolor}
\usepackage{soul}
\usepackage{physics}
\usepackage{graphics}
\usepackage[ruled,vlined]{algorithm2e}

\newlength{\commentWidth}
\newcommand{\atcp}[1]{\tcp*[r]{\makebox[\commentWidth]{#1\hfill}}}

\usepackage[colorlinks=true]{hyperref}

\usepackage{dcolumn}
\usepackage{bm}

\begin{document}

\title{Multi-Objective Bayesian Optimization for Online Accelerator Tuning}

\author{Ryan Roussel}%
\email{rroussel@uchicago.edu}
\affiliation{Department of Physics, University of Chicago, Chicago, Illinois 60637, USA}%
\author{Adi Hanuka}%
\affiliation{SLAC National Laboratory, Menlo Park, California 94025, USA}%
\author{Auralee Edelen}%
\affiliation{SLAC National Laboratory, Menlo Park, California 94025, USA}%

\date{\today}

\begin{abstract}
Particle accelerators require constant tuning during operation to meet beam quality, total charge and particle energy requirements for use in a wide variety of physics, chemistry and biology experiments. 
Maximizing the performance of an accelerator facility often necessitates multi-objective optimization, where operators must balance trade-offs between multiple objectives simultaneously, often using limited, temporally expensive beam observations. 
Usually, accelerator optimization problems are solved offline, prior to actual operation, with advanced beamline simulations and parallelized optimization methods (NSGA-II, Swarm Optimization).
Unfortunately, it is not feasible to use these methods for online multi-objective optimization, since beam measurements can only be done in a serial fashion, and these optimization methods require a large number of measurements to converge to a useful solution.
Here, we introduce a multi-objective Bayesian optimization scheme, which finds the full Pareto front of an accelerator optimization problem efficiently in a serialized manner and is thus a critical step towards practical online multi-objective optimization in accelerators.
This method uses a set of Gaussian process surrogate models, along with a multi-objective acquisition function, which reduces the number of observations needed to converge by at least an order of magnitude over current methods.
We demonstrate how this method can be modified to specifically solve optimization challenges posed by the tuning of accelerators.
This includes the addition of optimization constraints, objective preferences and costs related to changing accelerator parameters. 
\end{abstract}

\maketitle

\section{\label{sec:intro}Introduction}
Accelerator optimization during operation (i.e. ``online tuning'') is a tedious but often necessary part of any experimental facility's operation.
Due to their large number of components and the impact of variable external factors, such as vibrations or temperature changes, accelerators must be continuously re-tuned and optimized to meet various beam quality objectives.
This requires hours of dedicated beamline time where teams of operational experts diagnose issues with beam quality and make corrections, for even small to medium size facilities.
This severely limits beam time that is available to experimenters (i.e. "users") thus reducing the facilities' overall scientific output.
With current technological advances in the fields of computer science and machine learning, it would be beneficial to have an automated or semi-automated algorithm take care of normal beamline tuning, reducing downtime while also allowing human experts to tackle more challenging operational and design problems.

As a response to this problem, a number of algorithms have been used to optimize current accelerator facilities.
Gradient-based algorithms, such as robust conjugate direction search (RCDS) \cite{huang_algorithm_2013}, have been used successfully in the past to optimize beam parameters.
Heuristic methods such as the Nelder-Mead Simplex \cite{nelder_simplex_1965} algorithm can also be used to optimize black box problems, when functional derivative information is not easily accessible.
More recently, BOBYQA has also been used to optimize accelerators \cite{powell_bobyqa_2009,neveu_photoinjector_2017,appel_beam_2019} by fitting data to a second order model in a local trust region, which accounts for noisy observations.
However, these methods can struggle to handle problems with many local minima and must be restarted many times to ensure a global solution is found.


Bayesian optimization \cite{shahriari_taking_2016, greenhill_bayesian_2020} provides a framework for global optimization, while significantly reducing the number of physical observations needed to find solutions while also taking into account functional noise.
In this method, physical observations are combined with a kernel, which describes the overall functional behaviour, to create what is commonly referred to as a Gaussian Process (GP) model, which predicts the value and uncertainty of a target function \cite{rasmussen_gaussian_2006}.
Using this prediction, an optimizer can then choose input points that are likely to be ideal, before a physical measurement is made.
Recently, this method was successfully used to efficiently optimize single objective problems at LCLS and SPEAR3, with a lower number of observations needed than Nelder-Mead Simplex and RCDS algorithms \cite{hanuka_online_2019, mcintire_bayesian_2016, duris_bayesian_2019, kirschner_adaptive_2019}.

These algorithms have been used to optimize a single beam characteristic, while in reality, accelerator tuning generally seeks to simultaneously optimize multiple facets of the beam (``objectives'') at a time.
This presents an issue, as individual beam characteristics can often be optimized only at the expense of others.
For example, it is difficult to simultaneously minimize both the bunch length and the transverse emittance of a low energy electron beam in a photo-injector due to space charge forces \cite{neveu_parallel_2019}.
Solving multi-objective problems during simulated beamline design has recently become a relatively simple task, given the development of evolutionary algorithms and the availability of large computing clusters, which can run a large number of particle physics simulations in parallel \cite{li_genetic_2018,neveu_parallel_2019}.

By contrast, solving multi-objective optimization problems during accelerator operations presents an extremely difficult challenge due to several factors.
Most notably, accelerator operators can only evaluate or observe the objectives for a single set of input parameters at any time (referred to as ``serialized observations").
This makes the use of evolutionary algorithms practically infeasible, due to the number of observations needed to converge to a solution if used in a serialized manner. 
Furthermore, an online optimization algorithm must be able to keep track of constraining functions, as well as account for relative objective preferences specified by the operators.
Finally, the optimization algorithm should take into account the costs of changing accelerator parameters during optimization.

In this paper, we use the recent development of Multi-Objective Bayesian Optimization (MOBO) \cite{emmerich_multicriteria_2016} to extend online Bayesian optimization of accelerators to solving multiple objective problems using serialized observations.
We also demonstrate how to extend this algorithm to solve specific practical challenges associated with online accelerator optimization.

\section{Online Multi-Objective Optimization of Accelerators}
We start with a brief explanation of techniques currently used to solve multi-objective problems.
This serves to motivate use of the MOBO algorithm for online accelerator optimization.

A simplistic way of solving a multi-objective optimization problem is to explicitly weight each objective relative to one another $a\ priori$, and add up the weighted objective values \cite{scheinker_online_2020, cropp_preperation_nodate}.
This optimization method results in a solution found only for a single set of weights (trade-offs), and must be repeated from scratch to explore different trade-offs between objectives.
Mapping out the full set of optimal trade-offs in an accelerator is highly desirable, particularly at facilities which must accommodate a variety working points, or when operators wish to benchmark beam simulation results to experimental realities.
In this case, repeating the optimization for a discrete set of weights is relatively inefficient, even when Bayesian optimization methods are used \cite{zuhal_comparative_2019}.

\begin{figure}
	\centering
	\includegraphics[width=0.75\linewidth]{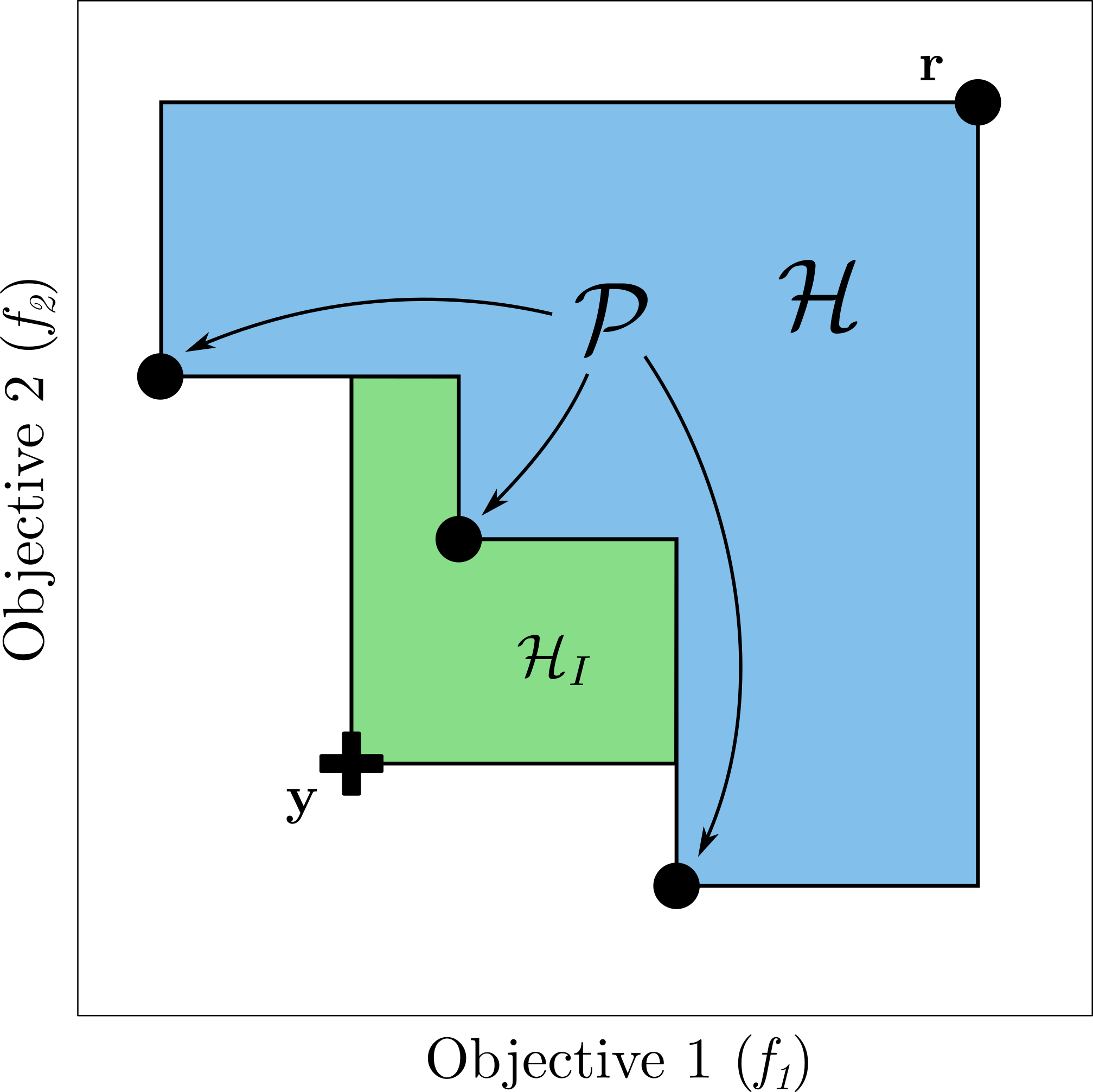}
	\caption{\label{fig:hypervolume_improvement} Cartoon of multi-objective optimization where each objective is to be minimized. Multi-objective optimization attempts to find a set of points known as the Pareto front $\mathcal{P}$ that dominate over a reference point $\mathbf{r}$ and any other observed points in objective space. The Pareto front hypervolume $\mathcal{H}$ (shown in blue) is the axis-aligned volume enclosed by the Pareto front and a reference point $\mathbf{r}$. Making a new observation $\mathbf{y}$, that dominates over points in the current Pareto front,  leads to an increase in hypervolume (shown in green), referred to as the hypervolume improvement $\mathcal{H}_I$.}
\end{figure}

To achieve this goal, multi-objective optimization algorithms attempt to find a set of points, known as the Pareto front $\mathcal{P}$, that optimally balances the trade-offs between multiple competing objectives simultaneously  (see Fig. \ref{fig:hypervolume_improvement}).
The Pareto front is defined as the set of ``non-dominated" points in objective space with respect to a reference point $\mathbf{r}$ (which itself must be dominated by every other observed point).
Points are non-dominated if they are as good as any other observed point for every objective and are better than any other point for at least one objective.
The hypervolume metric $\mathcal{H}$, \cite{naujoks_multi-objective_2005} shown in Fig. \ref{fig:hypervolume_improvement}, is often used to characterize the quality of the Pareto front, where a larger volume corresponds to a better solution set.
Adding observations to the current data set which dominate over points in the current Pareto front leads to an expansion of the hypervolume, characterized by the hypervolume improvement $\mathcal{H}_I$ (see Fig. \ref{fig:hypervolume_improvement}.
Algorithms generally stop once this metric converges to a maximum as new observations are added, signifying that a correct approximation to the true Pareto front has been reached.

A popular set of techniques used to solve multi-objective problems are known as evolutionary algorithms.
These algorithms, such as Non-dominated Sorting Genetic Algorithm II (NSGA-II) \cite{deb_fast_2002} or Multi-Objective Particle Swarm Optimization \cite{kennedy_particle_1995, huang_nonlinear_2014,pang_multi-objective_2014}, are based on the generation of a large collection of candidate solutions, which are then observed via simulation or experiment, usually in a parallelized manner.
The results from each observation are then sorted into non-dominated and dominated subsets.
The non-dominated subset of candidate solutions is used to produce the next ``generation" of candidate solutions using a stochastic heuristic, which are then re-evaluated.
The process is repeated over a number of generations until the non-dominated set of observations converges to a stationary Pareto front or the hypervolume has converged to a maximum value.
It has been shown that these methods are well suited for solving accelerator design optimization problems \cite{li_genetic_2018,neveu_parallel_2019}.

Recently, surrogate assisted evolutionary algorithms have been developed which combine evolutionary algorithms with surrogate models of the objective functions \cite{huang_multi-objective_2019, edelen_machine_2020}.
A fast executing surrogate model (possibly a GP or a neural network) is used to predict if a candidate solution generated by the evolutionary algorithm will be Pareto optimal before an observation is made.
This significantly speeds up convergence over basic evolutionary algorithms by eliminating candidate observations that are not predicted to improve the Pareto frontier.
The surrogate model is retrained after observations are made, thus improving the model's accuracy as the optimization progresses.
This further improves the convergence speed of the algorithm, as the surrogate model gains knowledge about the objective functions and can more adequately identify which candidates will be non-dominated.

However, these algorithms still are not ideal for online accelerator optimization.
Evolutionary type algorithms rely on parallelized evaluation of the objective functions, which is inefficient when restricted to serialized evaluations, as is the case for online accelerator optimization. 
Furthermore, evolutionary algorithms use a binary classification metric of Pareto dominance to identify which candidate points to observe.
As a result, this metric does not guarantee optimal expansion of the Pareto front, as it does not consider the relative hypervolume improvement of individuals in the non-dominated subset of candidates.
This further reduces the observation efficiency of these algorithms, in a case where efficiency is paramount.

The Multi-Objective Bayesian Optimization (MOBO) algorithm \cite{emmerich_multicriteria_2016} achieves maximum efficiency by using an explicit calculation of the hypervolume improvement as an acquisition function to expand the Pareto frontier.
The hypervolume improvement $\mathcal{H}_I$ is defined as the increase in Pareto front hypervolume by adding a new observation $\mathbf{y}$ (see Fig. \ref{fig:hypervolume_improvement}).
Each objective is modeled using a GP surrogate model which can be used to predict the hypervolume improvement as a function of input parameters. 
By maximizing the hypervolume improvement acquisition function, MOBO can determine a single point that maximally increases the Pareto front hypervolume at every step in a serialized manner, making it ideal for online accelerator optimization.



\section{\label{sec:MOBO} Multi-Objective Bayesian Optimization}
We begin the explanation of MOBO by first briefly going over single-objective Bayesian optimization.
This will aid our understanding of different components inside , many of which will be applied directly to the multi-objective case.
To maintain consistency with reference texts, we assume that single objective optimization aims to maximize the objective, while in the multi-objective case we wish to minimize each objective.
Simply multiplying any objective and its corresponding observed values by -1 allows us to switch the optimization goal from maximization to minimization or vice versa.

\subsection{\label{subsec:SOBO} Single Objective Bayesian Optimization}
The goal of our optimization strategy is to maximize the function $f(\mathbf{x})$ using as few observations of $f$ as possible inside the input domain $\mathbf{x} \in \mathcal{X}$ . 
Bayesian optimization uses two components to achieve this. 

The first component is the GP surrogate model.
A ``surrogate model'' in this case refers to a computationally cheap-to-evaluate predictive model, that acts as a predictive stand-in for any computationally expensive or difficult to measure system, and can be either a local or a global model. 
The GP surrogate produces both the predicted mean $\mu(\mathbf{x})$ and the corresponding uncertainty $\sigma(\mathbf{x})$ of a random function value at the point $\mathbf{x}$: $f(\mathbf{x})\sim\mathcal{GP}(\mu(\mathbf{x}),k(\mathbf{x},\mathbf{x}'))$ where $k(\mathbf{x},\mathbf{x}')$ is the covariance (kernel) function.
The kernel function represents how we expect our function to change between two points in input space, for example, we can specify how rapidly $f(\mathbf{x})$ changes as a function of the distance between two points $\mathbf{x}$ and $\mathbf{x}'$.
Given a set of observations $\mathcal{D}_N = \{(\mathbf{x}_1,y_1),(\mathbf{x}_2,y_2),\dots,(\mathbf{x}_N,y_N)\}$ where $y_n$ is the observed value which is assumed to include normally distributed noise $\epsilon$, $y_n = f(\mathbf{x_n}) + \epsilon$, and we can then calculate the predicted mean and variance anywhere in input space. 
Details on creating and using a GP surrogate model can be found in Appendix \ref{sec:regression} and in Ref. \cite{rasmussen_gaussian_2006}.

\begin{algorithm}[htp]
\SetAlgoLined
\SetKwInOut{Input}{input}
\SetKwInOut{Output}{output}
\Input{input domain $\mathcal{X}$, dataset $\mathcal{D}$, GP prior $\mathcal{M}_0 = \mathcal{GP}(\mu_0,k_0)$, acquisition function $\alpha$, noise $\epsilon$}
 \For{$i = 1,2,3,\dots$}{
  $x_i \gets$ argmin$_{x\in\mathcal{X}}\alpha(x|\mathcal{M}_{i-1}$) \atcp{optimize $\alpha$}
  $y_i \gets f(x_i) + \varepsilon$ \atcp{do observation}
  $\mathcal{M}_i \gets \mathcal{M}_{i-1}|(x_i,y_i)$ \atcp{update model}
 }
 
 \caption{Bayesian Optimization (BO)}
 \label{algo:BO}
\end{algorithm}

The second component of Bayesian optimization is an acquisition function $\alpha(\mathbf{x})$ which codifies how promising potential observation points are, based on mean and uncertainty predictions from the GP model.
To find the global optimum efficiently we wish to search regions of input space that either take advantage of previously observed extremum points (exploitation) or have a large amount of uncertainty (exploration).
We choose our acquisition function such that it is maximized at the point of most interest, one that properly balances the trade off between exploration and exploitation. 
We can then use a standard single-objective optimization algorithm to optimize the cheap-to-evaluate acquisition function to propose the next observation location, instead of directly optimizing the expensive-to-evaluate physical experiment.  
Two popular acquisition functions for Bayesian optimization are expected improvement (EI) and upper confidence bound (UCB) \cite{jones_efficient_1998,srinivas_gaussian_2010}.

\begin{figure*}
	\centering
	\includegraphics[width=1.0\linewidth]{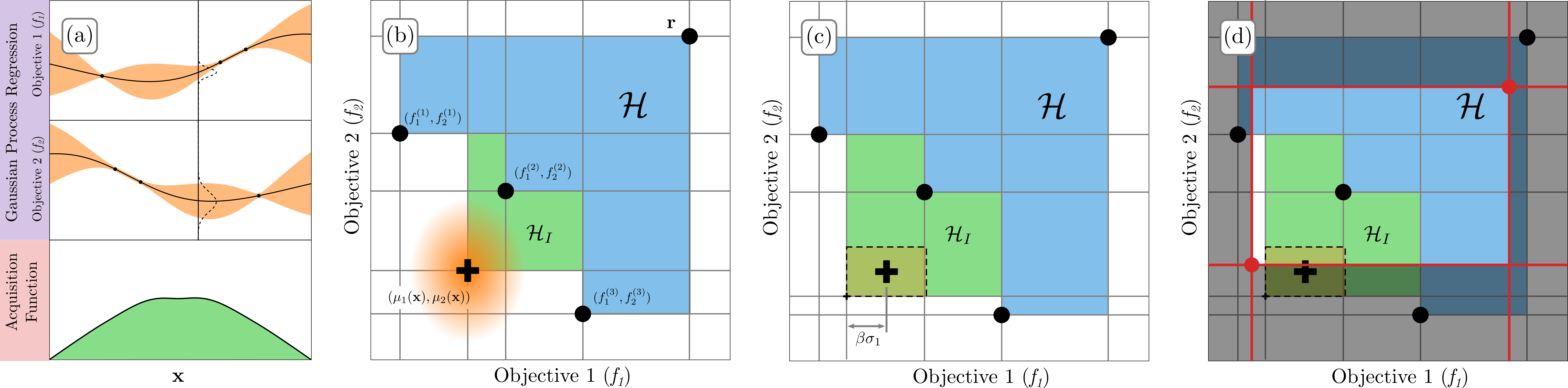}
	\caption{\label{fig:cartoon} Cartoons showing hypervolume improvement metrics used for MOBO. Blue regions denote current Pareto front hypervolume defined by a reference point (upper right) and three observed points. Green regions denote hypervolume improvement by adding an observation. (a) Each objective is modeled by an independent GP that predicts the function value and an uncertainty, the next optimization point is chosen by maximizing the acquisition function. (b) Expected hypervolume improvement (EHVI) where the distribution of predicted objective values is given by a probability distribution (orange shading) centered at the black cross. (c) Upper confidence bound hypervolume improvement (UCB-HVI) where the hypervolume improvement is determined by an optimistic view of the predicted objective values. (d) Truncated UCB-HVI where we only consider hypervolume contributions within the un-shaded sub-space $\mathcal{T}$.}
\end{figure*}

Expected improvement calculates the average improvement of a point over the best observed function value $f_{best}$
\begin{align}
\begin{split}
    \alpha_{EI}(\mathbf{x}) & = \mathbb{E}[\mathrm{max}(f_{best} - f(\mathbf{x}),0)]\\
                            & = (f_{best} - \mu(\mathbf{x}))\Phi\Big(\frac{f_{best} - \mu(\mathbf{x})}{\sigma(\mathbf{x})}\Big) \\&\ + \sigma(\mathbf{x}) \phi\Big(\frac{(f_{best} - \mu(\mathbf{x})}{\sigma(\mathbf{x})}\Big)
\end{split}
\end{align}
where $\Phi(\cdot)$ and $\phi(\cdot)$ are the probability distribution function and cumulative distribution function of a Gaussian distribution respectively. 

Upper confidence bound calculates the most optimistic improvement at a given point, weighted by a parameter $\beta$, which explicitly specifies the trade off between exploration and exploitation
\begin{equation}
    \alpha_{UCB}(\mathbf{x}) = \mu(\mathbf{x}) + \sqrt{\beta}\sigma(\mathbf{x}).
\end{equation}
For $\beta \ll 1$ UCB prioritizes exploitation; conversely for $\beta \gg 1$ UCB prioritizes exploration. 
This parameter can be increased as the optimization progresses, to prevent the optimizer from getting stuck in a local optimum \cite{srinivas_gaussian_2010}. 
Combining both the GP surrogate model and the acquisition function we can now perform Bayesian optimization using the method presented in Algorithm \ref{algo:BO}.

\subsection{\label{subsec:MOBO} Incorporating Multiple Objectives}
We now extend this methodology, following \cite{emmerich_multicriteria_2016}, to incorporate $P$ objectives $\mathbf{f} = \{f_1,f_2,\dots,f_P\}$. 
We assume that the objectives share the same input domain $x\in\mathcal{X}$ and are all observed for each input point such that $\mathcal{D}$ now contains the set of N observations of each objective, $\{(x_1,\mathbf{f}_1),(x_2,\mathbf{f}_2),\dots,(x_N,\mathbf{f}_N)\}$. 
Each objective is then modeled as an independent GP such that $f_p(\mathbf{x}) \sim \mathcal{GP}_p(\mu_p(\mathbf{x}),k_p(\mathbf{x},\mathbf{x'}))$, as seen in Fig. \ref{fig:cartoon}(a). 
Each GP has its own independent kernel which is trained on corresponding observations by maximizing the marginal log likelihood.

In order to proceed with optimization we must construct a scalar acquisition function $\alpha:\mathcal{X} \to \mathbb{R}$ that finds points which are likely to maximally increase the Pareto frontier hypervolume. 
We consider two acquisition functions that have been developed for this purpose. 

The first multi-objective acquisition function, expected hypervolume improvement (EHVI) seen in Fig. \ref{fig:cartoon}(b), is analogous to single-objective expected improvement.
This acquisition function calculates the average increase in hypervolume using the probability distribution of each objective function from the surrogate model. 
The EHVI acquisition function is formally defined as
\begin{equation}
    \alpha_{EHVI}(\mathbf{\mu},\mathbf{\sigma}, \mathcal{P},\mathbf{r}) := \int_{\mathbb{R}^P} 
    \mathcal{H}_I(\mathcal{P},\mathbf{y},\mathbf{r})\cdot \mathbf{\xi}_{\mathbf{\mu},\mathbf{\sigma}}(\mathbf{y})d\mathbf{y}
    \label{eqn:EHVI}
\end{equation}
where $\mathcal{P}$ is the current set of Pareto optimal points, $\mathbf{r}$ is the reference point, $\mathcal{H}_I(\mathcal{P},\mathbf{y},\mathbf{r})$ is the hypervolume improvement from an observed point $\mathbf{y}$ in objective space, and $\mathbf{\xi}_{\mathbb{\mu},\mathbf{\sigma}}$ is the multivariate Gaussian probability distribution function with the GP predicted mean $\mathbf{\mu}$ and standard deviation $\mathbf{\sigma}$ for each objective.
The hypervolume improvement is defined by the exclusive hypervolume contribution to the current Pareto front by adding $\mathbf{y}$ to the Pareto set, as seen in the green region in Fig. \ref{fig:cartoon}(b). 

The reference point $\mathbf{r}$ is chosen such that all expected observations $\mathbf{f}$ dominate the reference point.
Any predicted points in objective space that do not satisfy this condition will not contribute to the hypervolume improvement, and are thus never chosen as observation candidates.
It is also important to note for optimization purposes that the prior mean for each objective GP is set to the corresponding component of the reference point.
This ensures that in regions of input space where observations have not been made, the mean of the GP model returns to the reference point.
As a result, the acquisition function will never predict that unobserved regions of input space contribute to the hypervolume.

The integral in Eqn. \ref{eqn:EHVI} can become computationally expensive to calculate, as the objective space must be decomposed into cells for which the integral has an analytical form.
This can become computationally expensive for high dimensional objective spaces with even a small number of observations, as a naive decomposition algorithm scales as $\mathcal{O}(N^P)$.
Work in this field has produced efficient methods for objective space decomposition, which improves scaling to $\mathcal{O}(N\log N)$ for 2-3 dimensions and $\mathcal{O}(2^{P-1}N^{P/2})$ scaling when $P \geq 4$ \cite{yang_efficient_2019}.
Regardless, this computational complexity results in a significant increase in computation time when used to maximize the acquisition function, which could be called many times depending on the optimization strategy.

The second multi-objective acquisition function, Upper Confidence Bound Hypervolume Improvement (UCB-HVI), is similar to the UCB acquisition function for single objective optimization. This acquisition function describes an optimistic view of the hypervolume improvement given the surrogate model prediction,
\begin{equation}
    \alpha_{UCB-HVI}(\mathbf{\mu},\mathbf{\sigma},\mathcal{P},\mathbf{r},\beta):= \mathcal{H}_I(\mathcal{P}, \mathbf{\mu} - \sqrt{\beta}\mathbf{\sigma},\mathbf{r}).
\end{equation}
The simplicity of this acquisition function reduces computation time relative to EHVI, especially in high-dimensional objective spaces. 
This is due to the development of advanced hypervolume computation strategies, such as the Walking Fish Group (WFG) algorithm \cite{while_fast_2012} or approximate hypervolume computation algorithms \cite{tang_fast_2020}.
These algorithms can be used to calculate the hypervolume improvement by projecting points from the Pareto front onto the sub-domain that is dominated by the test point.
The use of these advanced algorithms allows UCB-HVI to be a much faster calculation than EHVI when the number of objectives is large ($P>3$), while still achieving similar optimization performance.
As a result we exclusively use the UCB-HVI acquisition function as a starting point to perform multi-objective optimization for the rest of the paper. 

We now show how the MOBO tackles a simple 2-objective optimization problem in 2D input space, $\mathbf{x} = (x_1,x_2)$. The problem is stated as 
\begin{align}
    \begin{split}
    \mathrm{minimize}\ & \{f_1(\mathbf{x}),f_2(\mathbf{x})\}\\
    f_1(\mathbf{x}) & = ||\mathbf{x} - \mathbf{1}||\\
    f_2(\mathbf{x}) & = ||\mathbf{x} + \mathbf{1}||\\
    x_n  & = [-2,2],\ n = 1,2.   
    \end{split}
    \label{eqn:toy_problem}
\end{align}
The analytical Pareto front for this problem in objective space lies on the line segment from $(f_1,f_2) = (0, 2\sqrt{2})$ to $(2\sqrt{2},0)$. 
In input space, Pareto optimal points lie on the line segment from $(x_1,x_2) = (-1,-1)$ to $(1,1)$. 
We start with a set of 5 random input points, drawn from a uniform distribution, which are used to initialize the GP surrogate model.
We choose an isotropic radial basis function (RBF) for our kernel (see Eqn. \ref{eqn:RBF}) with a length-scale of $\lambda = 1.0$ and a variance of $\sigma_f^2 = 0.5$ which roughly matches the functional form of our objectives.
The UCB-HVI acquisition function with $\beta = 0.01$ is then used to determine the next point to sample.
This value of $\beta$ is chosen to heavily weight exploitation since our functions are unimodal. 

\begin{figure}[htp]
	\centering
	\includegraphics[width=1.0\linewidth]{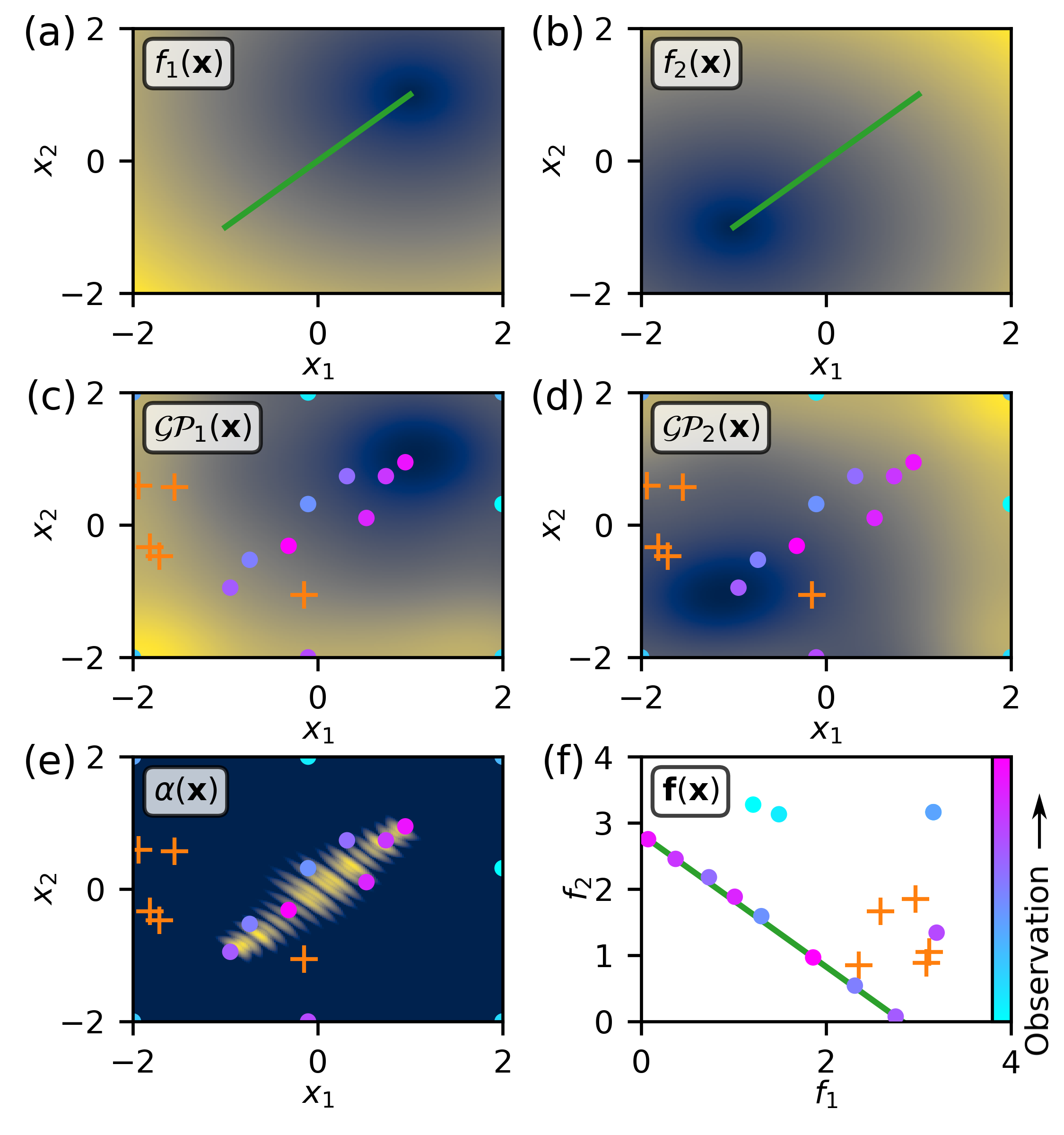}
	\caption{\label{fig:toy_mobo} Optimization results after 15 function observations using the MOBO algorithm on the problem defined in Eq. \ref{eqn:toy_problem}. Functions plotted in (a-e) are normalized to the range 0 (blue) to 1 (yellow). (a-b) Ground truth of $f_1,f_2$. Green line denotes the analytical location of Pareto optimal points. (c-d) GP mean prediction for $f_1,f_2$ respectively. (e) UCB-HVI acquisition function with $\beta = 0.01$. (f) Plot of observation values in objective space and the analytical Pareto front (green line). (c-f) Colored dots denote observation locations in input and output space, colored by observation number (blue to pink). Orange crosses denote the locations of 5 random, initial observation points.}
\end{figure}

Fig. \ref{fig:toy_mobo} shows optimization results after 15 optimization iterations.
We see that the GP prediction for each of the objectives (Fig. \ref{fig:toy_mobo}(c),(d)) near the observed points closely resembles the true value of each objective (Fig. \ref{fig:toy_mobo}(a),(b)).
After only 15 observations the Pareto front found by the UCB-HVI algorithm closely matches the analytical one seen in Fig. \ref{fig:toy_mobo}(f). 
We also observe that extrema of the acquisition function Fig. \ref{fig:toy_mobo}(e) (i.e. the most likely points to expand the Pareto front hypervolume) are located near the analytical Pareto optimal region in input space.
Each successive extremum is located in between each previously observed point due to the increase in uncertainty in-between observations.
This means that after the algorithm observes points along the entire Pareto front at a low resolution, the algorithm will continually attempt to increase the hypervolume by sampling intermediate points in-between previous observations.

\subsection{Adding optimization preferences and constraints}
One major advantage of the MOBO approach is the ability to specify how the optimizer searches the input space when considering preferential treatment of objectives and adding constraints. 
In the case of preferential treatment, we wish to specify that the optimizer searches in a given objective subspace (as seen in Fig. \ref{fig:cartoon}(d)), thus optimizing one objective or a set of objectives over another.
To achieve this, we simply set the acquisition function to zero outside of the selected sub-domain \cite{yang_truncated_2016}.
On the other hand, if we wish to specify an objective constraint, we require that an observed objective quantity $g(\mathbf{x})$ satisfies $g(\mathbf{x}) \leq h$ where $h$ is a constant.
In this case, we create a surrogate model for $g(\mathbf{x})$ and use it to predict the probability that the constraint will be satisfied \cite{gardner_bayesian_2014}. 
We then multiply the acquisition function by this probability to bias the optimizer against choosing points in a region that will likely violate the constraint.

While at first glance it seems that these two methods would result in the same behaviour, i.e. limiting the region where the acquisition function is non-zero, they in fact produce different results.
The addition of a preferential objective sub-domain results in a Pareto front that is only found within that sub-domain, which implies that all objectives are still minimized within the sub-domain (which comes at the expense of other objectives).
On the other hand, a constraint loosens this requirement, allowing any objective value that satisfies the constraint, which in-turn can lead to better solutions for the other objectives.
The subtle difference between these two methods allows more flexibility during optimization, suiting the different operation requirements for each accelerator. 
We now look at how to implement each of these methods in the MOBO framework.

The preferential algorithm specifies both a maximum and minimum reference point in objective space, as supposed to a single reference point in normal MOBO.
It then calculates what has been coined as the truncated hypervolume improvement \cite{yang_truncated_2016}. 
If we specify the truncated domain $\mathcal{T} \in [\mathbf{A},\mathbf{B}]$ defined by the minimum objective point $\mathbf{A}$ and the maximum objective point $\mathbf{B}$ the truncated expected hypervolume improvement (TEHVI) is thus given by
\begin{equation}
    \alpha_{TEHVI}(\mathbf{\mu},\mathbf{\sigma}, \mathcal{P},\mathbf{A},\mathbf{B}) := \int_{\mathbb{R}^P \in \mathcal{T}} 
    \mathcal{H}_I(\mathcal{P},\mathbf{y},\mathbf{r})\cdot \mathbf{\xi}_{\mathbb{\mu},\mathbf{\sigma}}(\mathbf{y})d\mathbf{y}
\end{equation}
where the Pareto set $\mathcal{P}$ is projected onto the truncated domain.
In a similar fashion, the truncated version of the UCB-HVI is given by
\begin{align}
    \begin{split}
    \alpha_{TUCB-HVI}(\mathbf{\mu},\mathbf{\sigma},\mathcal{P},\beta,\mathbf{A},\mathbf{B}):=\\
    \begin{cases}
        \mathcal{H}_I(\mathcal{P}, \mathbf{y},\mathbf{B}) & \mathbf{y} \in \mathcal{T}\\
        0 & \mathrm{otherwise}
    \end{cases}
    \end{split}
\end{align}
where $\mathbf{y} = \mathbf{\mu} - \sqrt{\beta}\mathbf{\sigma}$.

If we wish to specify an inequality constraint that needs to be satisfied, we create a GP surrogate model that predicts the probability of that constraint being satisfied and modify our acquisition function accordingly.
We assume that we have another observed quantity $g(\mathrm{x})$ that must satisfy $g(\mathrm{x}) \leq h$ whose value is stored in a dataset $\mathcal{D}_g$ and used in a GP to predict the probability distribution of $g(\mathrm{x})$.
The probability of a point $\mathbf{x}$ satisfying the constraint condition is then
\begin{equation}
    P_g(\mathbf{x}) := Pr[g(\mathbf{x}) \leq h] = \int_{-\infty}^h p(g(\mathbf{x})|\mathcal{D}_g)dg(\mathbf{x})
\end{equation}
which is simply a univariate Gaussian cumulative distribution function.
This probability can be adapted to suit a number of different type of constraints by modifying the limits of this integral.
Now we can define a new constrained version of the acquisition function $\hat{\alpha}(\mathbf{x})$ as
\begin{equation}
    \hat{\alpha}(\mathbf{x}) = \alpha(\mathbf{x})P_g(\mathbf{x}).
\end{equation}
Our acquisition function will be reduced anywhere we predict the constraint has a high probability of being violated and remain unchanged where there is a high probability that the constraint is satisfied.
Extra constraints can be easily added by multiplying the acquisition function by their respective probabilities of satisfying each additional constraint.

\begin{figure}[htp]
	\centering
	\includegraphics[width=1.0\linewidth]{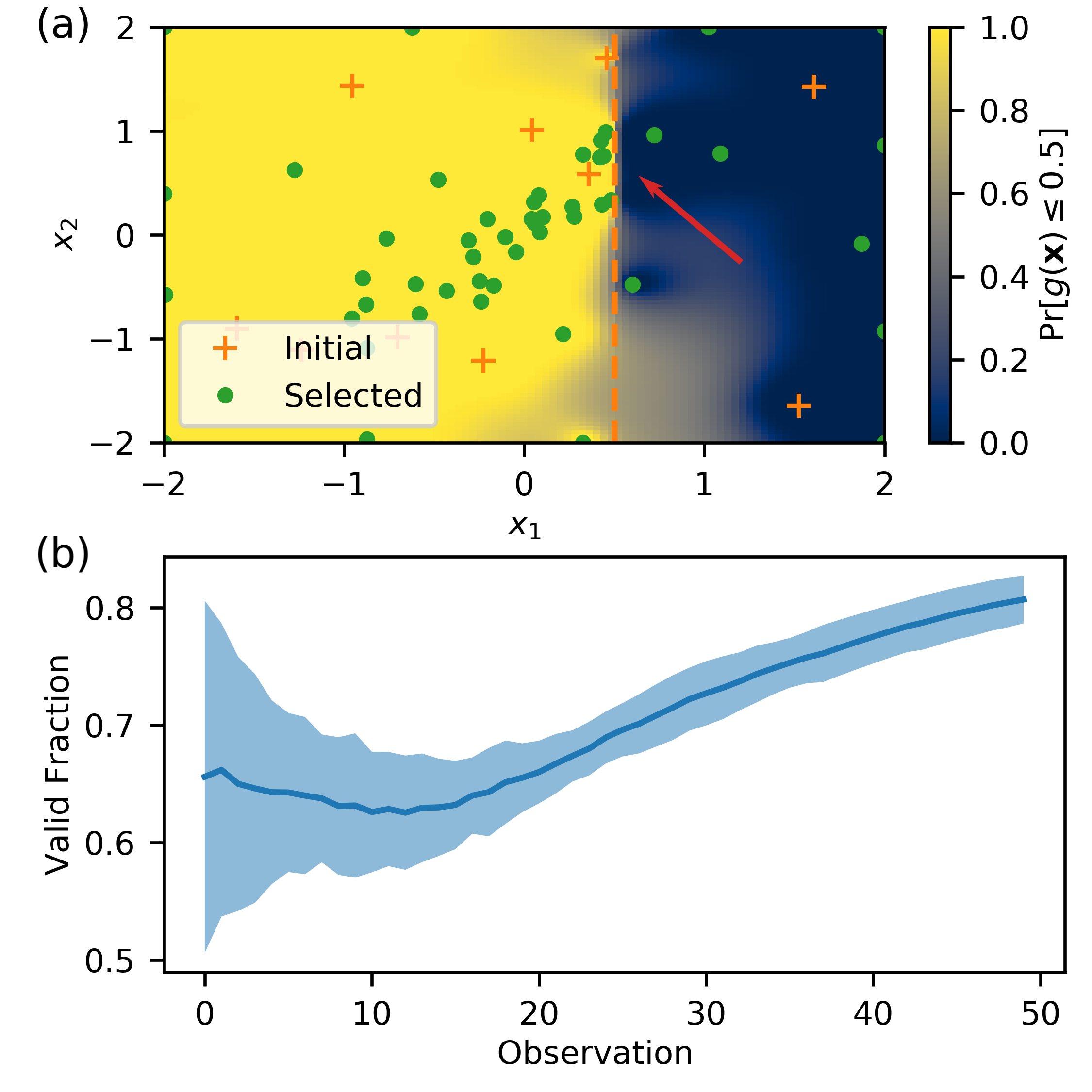}
	\caption{\label{fig:toy_constrained} (Color) (a) Probability map of satisfying the constraint $x_1 \leq 0.5$ after 50 observations, selected by constrained UCB-HVI, while optimizing Eq. \ref{eqn:toy_problem}. The red arrow points to where the constraint probability is most accurate, which is where the greatest number of observations are located. (b) Average fraction of points satisfying the constraint over 25 randomly initialized constrained MOBO optimization runs. Shading denotes one sigma spread.}
\end{figure}

An example of adding a constraint to the problem specified in Eq. \ref{eqn:toy_problem} is shown in Fig. \ref{fig:toy_constrained}. 
In this case we add the constraint inequality $g(\mathbf{x}) \leq 0.5$ where $g(\mathbf{x}) = x_1$, to stand in opposition of minimizing the first objective. 
We see the predicted probability that a point in parameter space will satisfy the constraint in Fig \ref{fig:toy_constrained}(a). 
Even though only a few observed points violate the constraint, we can clearly see a predicted threshold boundary, consistent with the stated constraint inequality. 
Furthermore, the boundary is best defined in the region of most interest, namely in the region denoted by the red arrow in Fig. \ref{fig:toy_constrained}(a) where Pareto optimal points would lie.
From Fig. \ref{fig:toy_constrained}(b) we see that as the optimization progresses the constraint function surrogate model accuracy is improved, and a smaller fraction of points which do not satisfy the constraint are sampled.
This leads to a reduction in the number of iterations needed to converge to a solution, as the constraint reduces the effective input parameter domain of the optimization problem.

\subsection{Proximal input space exploration}
One aspect of accelerator optimization that is often overlooked when constructing optimization algorithms is the cost associated with exploring the input parameter space.
Changes to input parameters (magnetic field strengths, cavity phases, etc.) often take time that scales proportionally to the magnitude of the change.
As a result, it is undesirable or infeasible to make large changes in machine input parameters frequently. 
Thus it is desirable to modify the acquisition function so that each optimization step travels a small distance in parameter space during optimization (i.e. "proximal exploration").

Achieving this is done by modifying the acquisition function to prioritize points in input space that are near the current or most recently observed parameter setting.
We modify our normal acquisition function by multiplying a multivariate Gaussian distribution, centered at the most recently observed point in input space $\mathbf{x}_0$ and has a precision matrix $\mathbf{\Lambda}$, to produce a proximal UCB-HVI (P-UCB-HVI) acquisition function given by
\begin{equation}
    \Tilde{\alpha}(\mathbf{x},\mathbf{x}_0) = \alpha(\mathbf{x}) \exp\Big[-\frac{1}{2}(\mathbf{x} - \mathbf{x}_0)^T\mathbf{\Lambda}(\mathbf{x} - \mathbf{x}_0)\Big].
    \label{eqn:restricted_acq}
\end{equation}
The precision matrix in this case specifies the cost associated with changing each input parameter, where larger elements of the matrix correspond to a higher cost associated with changing a given parameter.
The matrix can be specified prior to optimization or trained from multiple optimization runs (Bayesian optimization has been used to optimize similar hyper-parameters for neural network regression \cite{snoek_practical_2012}).
The addition of this extra term decreases the acquisition function far away from the most recently observed point.
With this modification we expect the MOBO algorithm to sample points along the Pareto front in input space, provided that the objectives are smoothly varying.
This is almost always the case in accelerator optimization problems with continuously variable input parameters.
Since the weighting function is non-zero throughout the input domain, large jumps to explore regions of parameter space with high uncertainty are still allowed, only if the acquisition function $\alpha(\mathbf{x})$ is large enough to overcome the travel distance penalty.
This is in contrast to simply reducing the UCB-HVI parameter $\beta$ to limit exploration, which prevents any such considerations.
By including the extra multi-variate Gaussian term, we maintain the optimization algorithm's ability to escape local extrema to explore regions of unobserved input space, while significantly reducing the frequency of large jumps.

\begin{figure}[htp]
	\centering
	\includegraphics[width=1.0\linewidth]{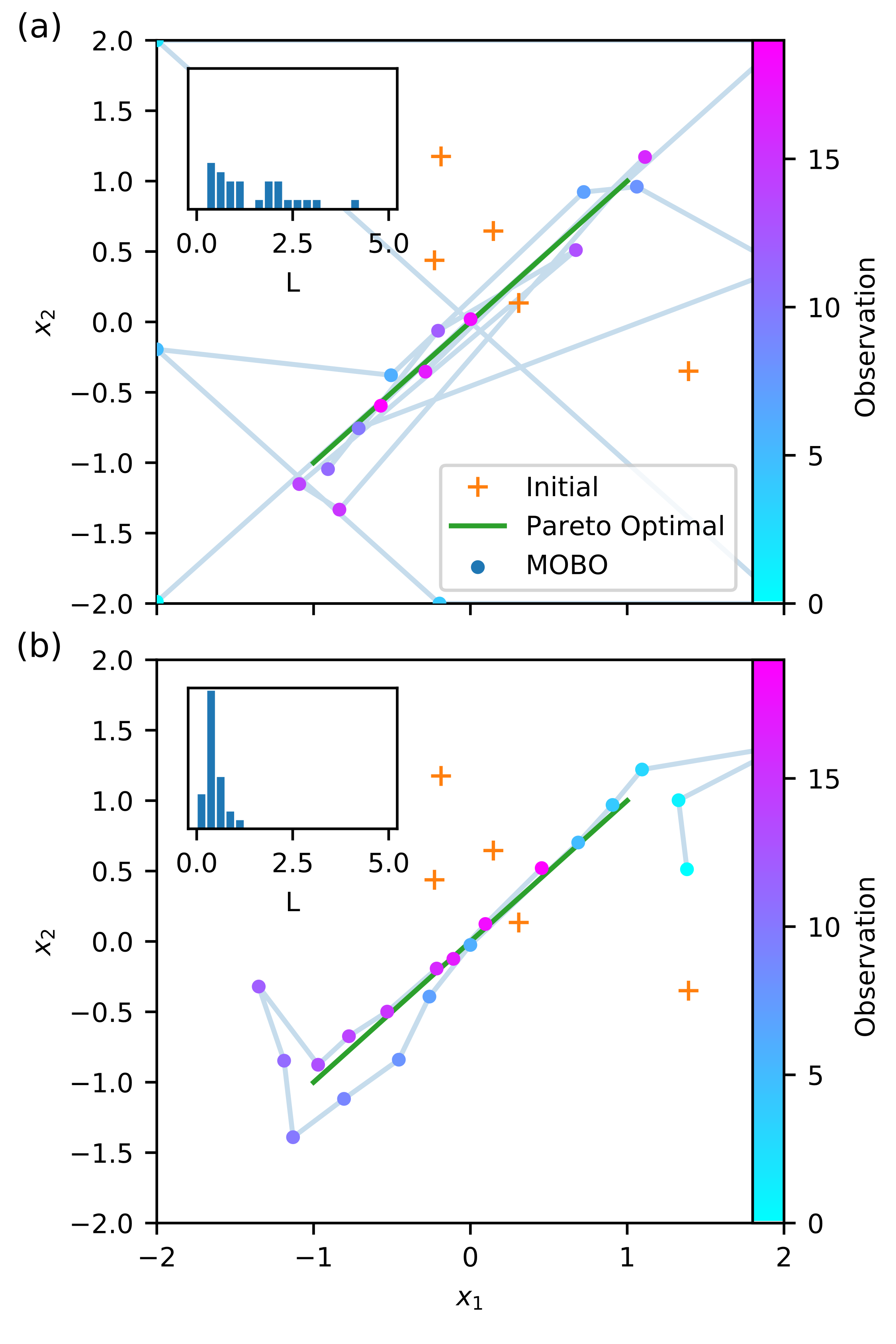}
	\caption{ (Color) Optimization trajectories in input space after 25 observations of objective functions in Eq. \ref{eqn:toy_problem} using (a) the unmodified UCB-HVI acquisition function and (b) the P-UCB-HVI acquisition function. Insets: Distribution of distances ($L$) traveled per step in input space.}
	\label{fig:traj}
\end{figure}

We demonstrate how this new method effects optimization by tracking how the UCB-HVI and P-UCB-HVI acquisition functions explore the input space, while optimizing our test problem Eq. \ref{eqn:toy_problem}. 
We start with a random sample of 5 points and then use UCB-HVI as our unmodified acquisition function to perform MOBO with 25 iterations, the result of which is shown in Fig. \ref{fig:traj}(a).
We then repeat the optimization with our modified acquisition function P-UCB-HVI.
We specify $\mathbf{\Sigma}= 4\mathbf{I}$ where $I$ is the identity matrix, the result of which is shown in Fig. \ref{fig:traj}(b).
The modified acquisition function reduces the average distance traveled in input space during each optimization step ($L$), when compared to UCB-HVI (see Fig. \ref{fig:traj} insets).
By tracking the order of observation, we see that while normal UCB-HVI seems to quasi-randomly explore the input space, P-UCB-HVI explores the Pareto optimal region in a disciplined manner.
It first travels along the Pareto optimal space until it reaches the end, then it explores in the vicinity of the endpoint to verify it is indeed the end of the Pareto optimal region.
Finally, it reverses direction and continues exploring the Pareto optimal space, jumping over regions that have already been explored.

\section{\label{sec:accelerator} Application to Accelerator Optimization}
\begin{figure}[htp]
	\centering
	\includegraphics[width=1.0\linewidth]{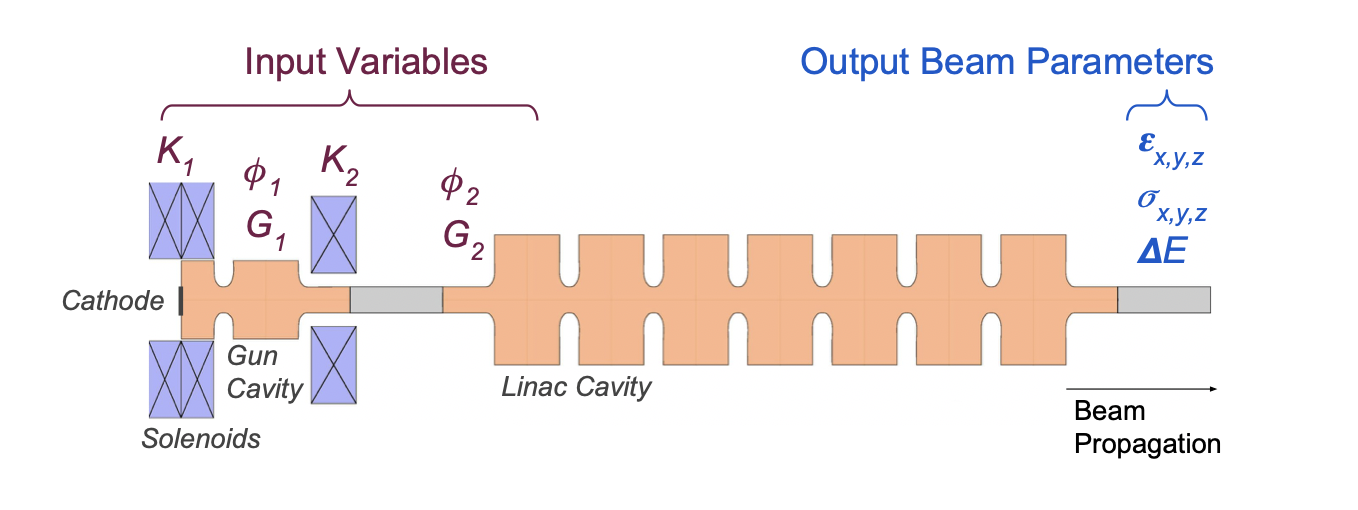}
	\caption{\label{fig:beamline_cartoon} Cartoon of the AWA photoinjector and first linac cavity. Input and output parameters used in optimization are labeled. Reproduced from \cite{edelen_machine_2020}.}
\end{figure}

We now demonstrate the MOBO framework on a multi-objective accelerator optimization problem, namely, optimizing the parameters of the Argonne Wakefield Accelerator (AWA) photoinjector \cite{conde_research_2017}.
The AWA photoinjector uses a Cesium-Telluride cathode in a copper, radio-frequency cavity, to produce electron beams with a wide variety of bunch charges for the use in wakefield accelerator physics experiments.
We consider a case where we can vary a number of parameters inside the injector and the first linac section, seen in Table \ref{TB:AWA_params} and in Fig. \ref{fig:beamline_cartoon}.
Our goal is to simultaneously minimize a collection of beam parameters at the exit of the linac section, also seen in Fig. \ref{fig:beamline_cartoon}.

\begin {table}[htp]
\caption{AWA Input Parameters}
\label{TB:AWA_params}
\begin{ruledtabular}
\begin {tabular}{l l l l l}
Parameter & Abbreviation & \begin{tabular}{@{}c@{}} Minimum\\value \end{tabular} &\begin{tabular}{@{}c@{}} Maximum\\value \end{tabular} & Unit \\
\colrule
\begin{tabular}{@{}l@{}} Solenoid 1\\\quad strength\end{tabular} & $K_1$ & 400 & 550 & $\mathrm{m^{-1}}$\\
\begin{tabular}{@{}l@{}} Solenoid 2\\\quad strength\end{tabular} & $K_2$ & 180 & 280 & $\mathrm{m^{-1}}$\\
Injector phase & $\phi_1$ & -10 & 0 & deg\\
Cavity phase & $\phi_2$ & -10 & 0 & deg\\
\begin{tabular}{@{}l@{}} Injector\\\quad accelerating \\\quad gradient\end{tabular}  & $G_1$ & 60 & 75 & MV/m\\
\begin{tabular}{@{}l@{}} Cavity\\\quad accelerating \\\quad gradient\end{tabular}  &  $G_2$ & 15 & 25 & MV/m\\
\end {tabular}
\end{ruledtabular}
\end{table}  

This problem was chosen based on previous work done towards creating a surrogate model of the AWA photoinjector \cite{edelen_machine_2020}.
In this previous work, the authors used the full 3D space charge, particle in cell (PIC) code OPAL \cite{adelmann_object_2009} to simulate a large set of randomly generated input parameters and measure the corresponding beam parameters at the injector exit.
They then created a neural network based surrogate model, trained on the simulation data set.
The model can be rapidly queried to retrieve output beam parameters for a given input parameter set.
The authors then showed that this surrogate model accurately reproduces results from the 3D PIC simulation.
We use this surrogate model for testing our optimization algorithm as it reduces simulation time by several orders of magnitude.
\begin{figure}[htp]
	\centering
	\includegraphics[width=1.0\linewidth]{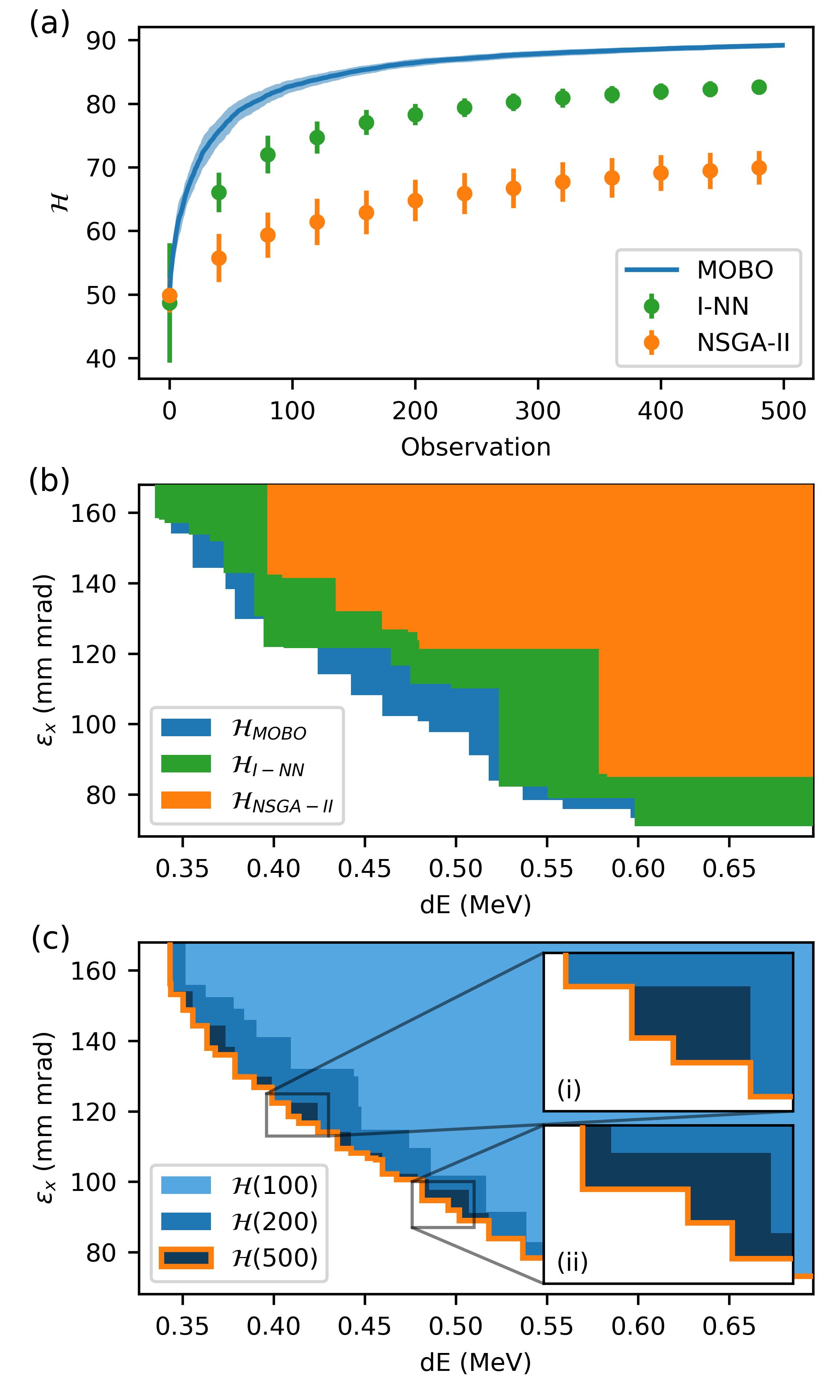}
	\caption{\label{fig:convergence} (Color) (a) Average Pareto front hypervolume $\mathcal{H}$ of 10 multi-objective optimization runs of the AWA example using MOBO, NSGA-II and iterated neural network (I-NN) algorithms. Shading and error bars denote 1 sigma variance. (b) Projected hypervolume onto energy spread ($dE$) vs. transverse emittance ($\varepsilon_x$) sub-space after 200 observations for each optimization algorithm shown in (a).  (c) Projected hypervolume onto $dE$ vs. $\varepsilon_x$ subspace after 100 (light blue), 200 (blue) and 500 (dark blue with orange outline) observations using the MOBO algorithm. Inset zoom (i) shows an increase in hypervolume due to increasing the Pareto front resolution, while inset zoom (ii) shows an increase in hypervolume due to finding new points that completely dominate old observations in projected space.}
\end{figure}

\subsection{Convergence Comparison}
\label{subsec:convergence}
In our first experiment, we use MOBO to minimize all seven exit beam parameters as a function of all six input parameters.
We wish to compare the convergence rate of MOBO with the convergence rates of standard and recently developed algorithms for solving multi-objective optimization problems.     
To begin with, all of the input and output values are normalized to the range $[-1,1]$ in order to account for the widely varying scaling of each parameter.
We assume that the functional form of each objective is smooth,and thus we choose the standard radial basis function kernel (Eq. \ref{eqn:RBF}) with an anisotropic precision matrix $\Sigma = \mathrm{diag}(\mathbf{l})^{-2}$ where $\mathbf{l}$ is a vector that stores an independent length-scale for each input parameter.
Initially, a randomly generated Latin-Hypercube distribution of 20 input parameter sets with corresponding objective observations is used to train each objective GP.
Hyperparameter training is done by maximizing the log marginal likelihood of the GP model with the gradient based Adam optimization algorithm \cite{kingma_adam_2017}, with 5000 iterations and a learning rate of 0.001.

Once trained, we use the UCB-HVI acquisition function to perform multi-objective Bayesian optimization with 500 sequential observations.
In this case, empirical testing found that $\beta=0.01$ gave the fastest convergence, likely due to the unimodal nature of each objective function, which allows us to aggressively exploit the GP model for the global extremum without worrying about getting caught in local extrema.

Maximizing the acquisition function is done via particle swarm optimization, implemented using the PyGMO package \cite{biscani_parallel_2020} with 64 individuals over 10 generations.
In order to account for new information gained from observations during optimization, we retrain the GP kernel hyperparameters with the accumulated data set every 10 observations, again using the Adam algorithm with a learning rate of 0.001 but with 1000 steps.

We re-run this optimization procedure 10 times, each with a different set of 20 randomly generated initial points every time.
After each observation, we calculate the exact hypervolume of the Pareto set in normalized objective space, referenced to the maximum possible value for each objective (in this case 1).
The average and variance of the hypervolume as a function of observation number after ten optimization runs is shown in Fig. \ref{fig:convergence}.

For comparison we run the same optimization test, but with previously used methods for multi-objective optimization, evaluated in serial, as would be the case during online optimization.
The first, Non-dominated Sorting Genetic Algorithm II (NSGA-II) \cite{deb_fast_2002}, is a popular genetic optimization algorithm, which has been previously used to solve multi-objective accelerator design problems \cite{neveu_parallel_2019}.
We conducted 10 optimization runs using the NSGA-II algorithm, with a population of 20 individuals, that matched the input parameter sets used in the MOBO optimization runs.
We then evolved the population for 200 generations, with a crossover probability of 0.8 and mutation probability of 0.05.
The hypervolume after the first 25 generations (500 function observations) is shown in Fig. \ref{fig:convergence}(a).

The second algorithm, iterated neural network (I-NN) optimization  \cite{edelen_machine_2020}, is a recently developed algorithm using surrogate neural network (NN) models to choose future observation locations.
In this method, observations are used to train a NN surrogate model, which in turn is optimized by the NSGA-II algorithm to propose a new set of observations that are likely to be non-dominated.
We plot the predicted hypervolume from the NN surrogate model after each batch of measurements in Fig. \ref{fig:convergence}(a).

From this comparison, we clearly see that MOBO reaches convergence much faster than both the NSGA-II and I-NN algorithms.
While not shown in Fig. \ref{fig:convergence}(a) it took about 17,500 NSGA-II observations to reach the same hypervolume that MOBO reached after 500 observations, roughly a factor of 35 times slower.

In Fig. \ref{fig:convergence}(b) we show the 2D projected Pareto front on the energy spread $dE$ and horizontal beam emittance $\varepsilon_x$ objective space from each of these optimization algorithms after 200 observations each. 
We observe that the Pareto front generated by NSGA-II is far from optimal and contains few points.
This is a direct result from NSGA-II's inefficient sampling behavior.
The points that NSGA-II chooses to observe are frequently dominated by previous observations due to the randomized heuristic used to generate observation proposals.
The I-NN algorithm improves over NSGA-II by including a neural network surrogate model that directs NSGA-II towards observing non-dominated points.
Neither of these algorithms include a direct calculation of the hypervolume increase for each proposed observation point, and thus do not optimally increase the hypervolume at each observation step.
As a result the Pareto front generated by MOBO is larger and has a higher resolution than the Pareto fronts generated by either NSGA-II or I-NN.
Generally, MOBO shows similar improvements in optimization speed when used in solving a variety of different optimisation problems with varying input and objective spaces \cite{zhan_expected_2020,allmendinger_surrogate-assisted_2017}. 

In Fig. \ref{fig:convergence}(c) we show the 2D projected Pareto fronts generated by MOBO after 100, 200 and 500 observations.
Since the objective space is high dimensional, it takes a large number of observations (100-200) for the algorithm to build up a well-defined Pareto front.
Once the front is loosely meshed, the acquisition function can increase the hypervolume in one of two ways, by either finding points in objective space in between prior observations, in order to improve the Pareto front resolution (Fig. \ref{fig:convergence}(c)(i)), or finding points in objective space that dominate initial observations (Fig. \ref{fig:convergence}(c)(ii)).
We see that most of the points present in the Pareto front after 200 observations remain on the Pareto front after 500 observations, as the majority of new observations lie in between prior observations in objective space.
We conclude that in this optimization run, the algorithm correctly predicts where the true Pareto front lies most of the time, leading to a gain in optimization efficiency.
This is in contrast to the heuristic methods we compare MOBO with, where only a small fraction of the observed points are actually on the true Pareto front.


\subsection{Constrained optimization}
\begin{figure}
	\centering
	\includegraphics[width=1.0\linewidth]{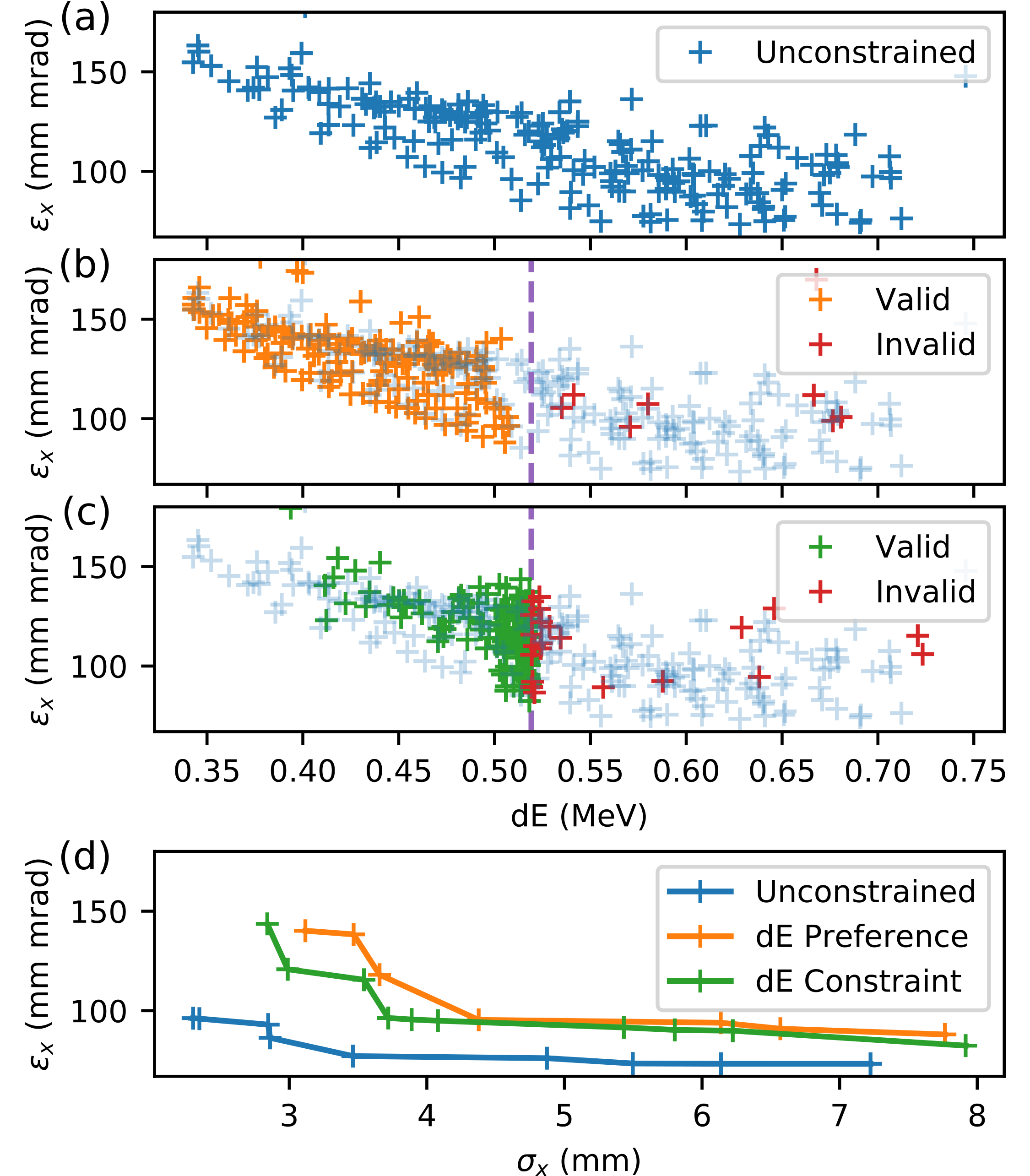}
	\caption{\label{fig:constrain_comparsion} (Color) Plots showing energy spread ($dE$) and horizontal beam emittance ($\varepsilon_x$) observations after 300 observations taken by MOBO algorithms. (a) MOBO with no constraints. (b) MOBO with an optimization preference of $dE<0.52$ MeV. (c) MOBO with an inequality constraint of $dE < 0.52$ MeV. (d) Projected Pareto fronts for $\sigma_x$ vs. $\varepsilon_x$ for each case above. The dotted lines in (b) and (c) denotes the preference/constraint limit.}
\end{figure}

We now investigate the effect of preferential or constrained treatment of an objective on the accelerator optimization.
First, we consider a case where we want to optimize the same objectives as the previous problem but wish to only find solutions where the energy spread satisfies $dE < 0.52$ MeV ($dE < -0.25$ in normalized coordinates).
To judge how this modification effects optimization, we compare the observed points in the 2D energy spread $dE$ and horizontal emittance $\varepsilon_x$ objective space, after 200 iterations in Fig. \ref{fig:constrain_comparsion}.
We see in Fig. \ref{fig:constrain_comparsion}(a) projected observations when no constraints or preferences are added.
It is important to note that a large majority of the observations plotted here are on the 7D Pareto front, even though only a small fraction make up the projected front in 2D space. 
When preferential treatment is added to the acquisition function (Fig.\ref{fig:constrain_comparsion}(b)), the algorithm observes almost no points that violate this preference.
Furthermore, since the effective volume of the objective space is significantly reduced, the optimizer finds a higher quality 2D Pareto front in the same number of steps as the unconstrained case.

Second, we consider a case where we relax this preference, removing the energy spread minimization objective entirely and replace it with the inequality constraint $dE < 0.52$ MeV.
The resulting distribution of observations appears significantly different in this case (Fig. \ref{fig:constrain_comparsion}(c)).
In this case, more observations are made that violate the constraint than in the preferential treatment, which is necessary to accurately model the constraining function near the boundary.
Furthermore, the optimizer allows the energy spread to increase up to the constraint value, instead of attempting to minimize it, in order to better optimize the six remaining objectives.
We see this effect in Fig. \ref{fig:constrain_comparsion}(d) where the Pareto front for a different set of objectives, $\sigma_x$ and $\varepsilon_x$, is better when the energy spread preference is relaxed to a constraint.
The 2D front then improves again when the constraint is completely removed, owing to the fact that $dE$ is allowed to increase further when all constraints are removed.

\begin{figure}
	\centering
	\includegraphics[width=1.0\linewidth]{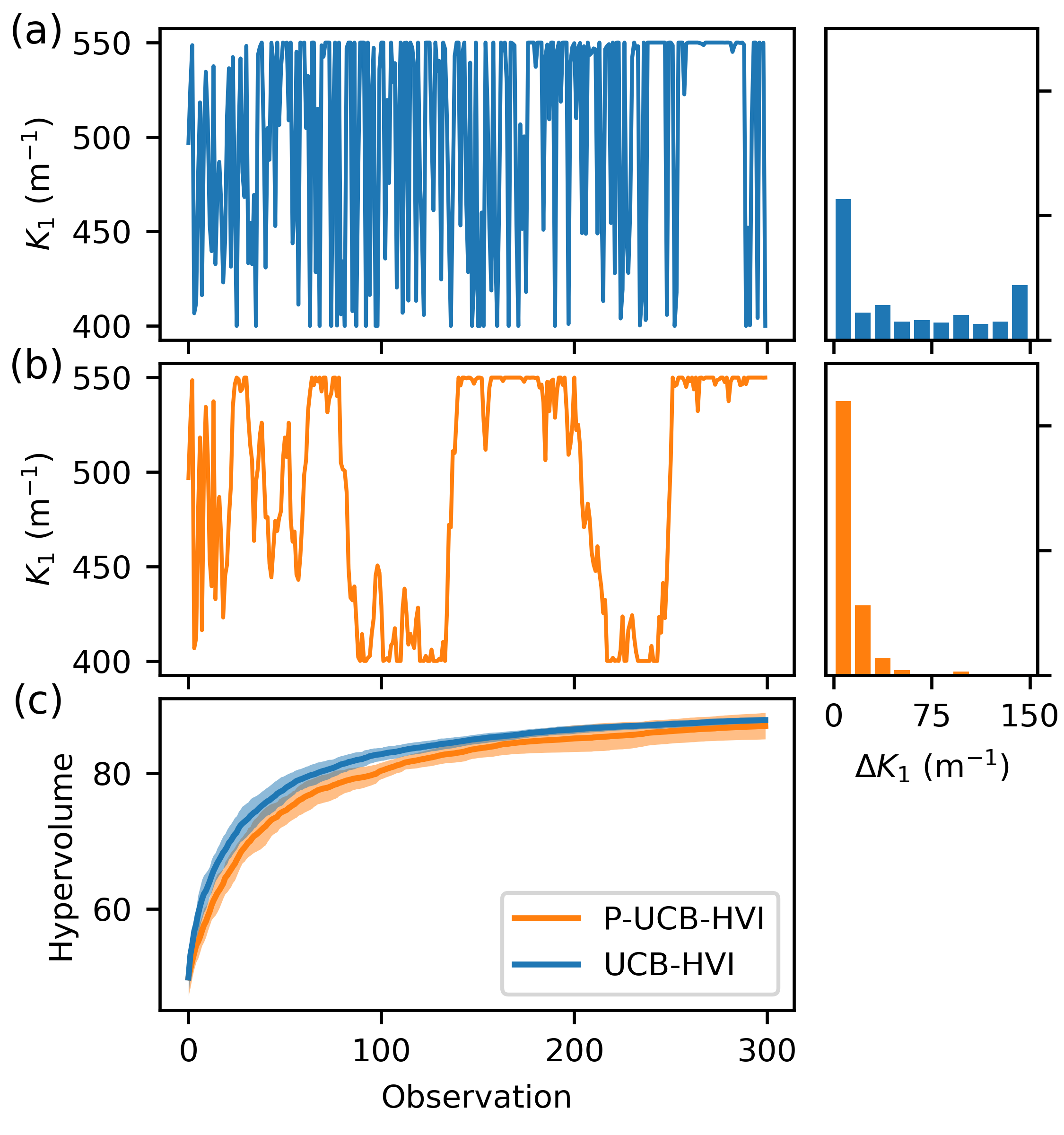}
	\caption{\label{fig:localized_comparsion} (Color) Comparison between normal UCB-HVI and the proximal UCB-HVI acquisition functions when used to perform optimization of the AWA photoinjector. (a) Solenoid 1 strength parameter over 300 observations using UCB-HVI (left) and corresponding distribution of $\Delta K_1$ for each step (right).
	(b) Solenoid 1 strength parameter and travel distance distribution when P-UCB-HVI is used. (c) Average Pareto front hypervolume of 10 optimization runs for UCB-HVI and P-UCB-HVI with identical random initialization sets. Shading denotes one sigma variance. }
\end{figure}
\subsection{Proximal optimization}
Finally, we demonstrate the use of P-UCB-HVI on optimizing the AWA problem.
We start with the same set of 10 initial sets of observations as in Section \ref{subsec:convergence} with the same hyperparameter training schedule.
However this time we run MOBO optimization using the P-UCB-HVI acquisition function, with an isotropic precision matrix (see Eq. \ref{eqn:restricted_acq}) $\mathbf{\Sigma} = 0.25\mathbf{I}$ is defined in normalized input space.

Results from these optimization runs are presented in Fig. \ref{fig:localized_comparsion}. 
We observe that during optimization, when the UCB-HVI acquisition function is used, the solenoid strength parameter $K_1$ is wildly varied to increase the hypervolume as much as possible each step.
However, when the localization term is added to the acquisition function, the frequency and amplitude of large jumps in parameter space are both decreased.
While not shown here, this change in behavior is mirrored in each of the other 5 input parameters.
Furthermore, the use of P-UCB-HVI acquisition function over the generic UCB-HVI function only minimally reduces the overall speed at which the method maximizes the Pareto front hypervolume (Fig. \ref{fig:localized_comparsion}c).

\section{\label{sec:conclusion} Conclusion}
In this paper we have demonstrated that the MOBO framework can be used to solve online multi-objective optimization accelerator physics problems.
This method efficiently finds the Pareto front in a serialized manner, which makes it viable for use in online accelerator optimization.
The framework also allows the user to easily specify objective preferences and constrain the objective space through the use of GPs.
Finally, we demonstrated that adding a localization term to the acquisition function effectively restricts the MOBO algorithm to prioritize moving through input space in a proximal, disciplined manner, which is especially important for practical use in accelerator facilities.

Our results also demonstrate the reasons why MOBO is ideal for online multi-objective accelerator optimization.
As stated previously, optimization takes place after every observation step, as opposed to methods designed for parallel use which are not sample efficient when used in a serialized context.
Second, observation points proposed by MOBO directly incorporate learned information about the objective function instead of using a heuristic method that is model independent to generate potential solutions.
As a result, MOBO strategically proposes solutions which optimally increase Pareto front hypervolume and improve Pareto front resolution.
While this comes at an increase in computational complexity, the extra computation time needed (estimated to be $<$ 5 s for most problems) is small relative to the reduction in optimization time associated with faster convergence to the Pareto front.

The use case for practical online multi-objective optimization can be viewed as experimental beamline characterization.
Accelerator working points are often predetermined through conducting multi-objective optimization on simulated versions of the beamline.
However, simulations rarely capture the full behavior of the beam in the real accelerator and as a result any selected working point from simulation might not be ideal in reality.
MOBO can be used to find the realistic ideal trade off between given objectives and the input parameters that are Pareto optimal during beamline commissioning or machine study.
Once a trade-off has been selected by specifying an explicit weighting of the objectives and the correct input parameters have been determined, single objective optimization strategies can take over during regular operation to do local optimization in response to noise and/or temporal drift.
Furthermore, the experimental multi-objective optimization process can be repeated periodically to monitor changes in machine performance over time and benchmark simulated Pareto fronts to experimental results.

While the goal of this work is to apply the MOBO framework towards online optimization of accelerators, this technique can also be applied to solve computational accelerator physics optimization problems.
High fidelity simulations of particle accelerator physics (beamlines, cavities, magnets etc.) are also a resource intensive process, often requiring time on computational clusters which have limited availability.
The recently developed $q$-Expected Hypervolume Improvement (qEHVI) \cite{daulton_differentiable_2020} algorithm is a parallelized extension of EHVI.
In contrast to EHVI, which proposes a single point at each optimization step, qEHVI proposes multiple $q$ points that are likely to increase the Pareto front hypervolume for each optimization step.
These points can be evaluated in a batched parallel process on a computing cluster, which significantly reduces overall optimization time while maintaining the sampling efficiency advantages of EHVI.
Furthermore, multi-fidelity approaches to MOBO have also been proposed to reduce optimization time \cite{belakaria_multi-fidelity_2020}, by incorporating low-cost, approximate simulations as a temporary stand-in for expensive high fidelity simulations.
The application of these methods to solving computational accelerator problems has the potential to dramatically reduce resource requirements for those in the field. 

\section{Acknowledgements}
This work was supported by the U.S. National Science Foundation under Award No. PHY-1549132, the Center for Bright Beams.

\appendix
\section{Gaussian Process Regression}
\label{sec:regression}
A GP regression model works by representing the function value at a given input point via a random variable drawn from a multivariate Gaussian distribution $f(\mathbf{x}) \sim \mathcal{GP}(\mu(\mathbf{x}), k(\mathbf{x},\mathbf{x'}))$ where $\mu(\mathbf{x})$ is the mean and $k(\mathbf{x},\mathbf{x'})$ is the covariance \cite{rasmussen_gaussian_2006}.
We start with a prior belief that $\mu(\mathbf{x})=0$ (without loss of generality) and use Bayes rule to update this belief to a new one (known as the posterior), conditioned on the observed data set $\mathcal{D}_N$ and the covariance function.
The prediction of the function at a test point $\mathbf{x}_*$ is given by $f_* = f(\mathbf{x}_*)$.
This random variable is then drawn from the conditional Gaussian 
\begin{align}
    f^*|\mathcal{D} &\sim \mathcal{N}(\mu_*,\sigma^2_*)\\
    \mu_* &= \mathbf{K}(\mathbf{x}_*,\mathbf{x}) \mathbf{K}(\mathbf{x},\mathbf{x})^{-1}\mathbf{y}\\
    \sigma^2_* &= \mathbf{K}(\mathbf{x}_*,\mathbf{x}_*) - \mathbf{K}(\mathbf{x}_*,\mathbf{x}) \mathbf{K}(\mathbf{x},\mathbf{x})^{-1} \mathbf{K}(\mathbf{x},\mathbf{x}_*)
\end{align}
where $\mathbf{K}(\mathbf{x}_*,\mathbf{x})$ is a $N\times1$ matrix that describes the covariance between the prediction point and the observed points, $\mathbf{K}(\mathbf{x},\mathbf{x})$ is a $N \times N$ matrix that describes the covariance between each pair of observation points and so on. 

The covariance function $k(\mathbf{x},\mathbf{x'})=\mathbf{K}(\mathbf{x},\mathbf{x'})$ is specified based on prior knowledge of the target functions' behavior. A common kernel is the radial-basis function (RBF) given by
\begin{equation}
    k_{RBF}(\mathbf{x},\mathbf{x'}) = \sigma_f^2 \exp\Big[-\frac{1}{2}(\mathbf{x} - \mathbf{x'})^T\mathbf{\Lambda}(\mathbf{x} - \mathbf{x'})\Big]
    \label{eqn:RBF}
\end{equation}
where $\Lambda$ is known as the precision matrix. In the isotropic case, the matrix is specified by $\mathbf{\Lambda} = \mathbf{I}/\lambda^2$ where $I$ is the identity matrix and $\lambda$ is referred to as the length-scale hyperparameter. This hyperparameter describes the characteristic length scale at which the function varies. For large $\lambda$ the function is expected to be smooth; decreasing $\lambda$ causes the function to vary quickly over a short distance. In a case where the function is expected to vary at different length scales along each input axis we can define an anisotropic kernel, were the matrix is specified by a length scale vector $\mathbf{l}$ where $\mathbf{\Lambda} = \mathrm{diag}(\mathbf{l})^{-2}$ where the diagonal elements specify a length scale for each input dimension. Further, the entire precision matrix can be specified via $\mathbf{\Lambda} = \mathbf{L}\mathbf{L}^T + \mathrm{diag}(\mathbf{l})$ where $\mathbf{L}$ is an upper triangle matrix to ensure that $\mathbf{\Lambda}$ is positive self-definite. Generally hyperparameters can be trained by maximizing the marginal log likelihood, which optimizes them to best fit the observed data set while minimizing the regression's functional complexity \cite{rasmussen_gaussian_2006}.
However, they can also be determined in a localized region by calculating the function's Hessian matrix at a given point \cite{hanuka_online_2019}.

\section{Code availability}
This research used only open source Python software libraries, including \texttt{GPFlow} \cite{matthews_gpflow_2017} and \texttt{Tensorflow} \cite{abadi_tensorflow_2016}. The algorithms developed are contained in a repository available at \url{https://github.com/roussel-ryan/Accelerator_MOBO}. The surrogate model of the AWA photo-injector is available upon request.

\bibliographystyle{unsrt}
\bibliography{main.bib}

\end{document}